\documentstyle[12pt,epsf]{article}
\begin{document}
\newcommand {\beq}{\begin{equation}}
\newcommand {\eeq}{\end{equation}}
\newcommand {\beqar}{\begin{eqnarray}}
\newcommand {\eeqar}{\end{eqnarray}}
\newcommand {\kagome}{Kagom\'e }
\begin{center}
{ \large \bf A hidden Goldstone mechanism in the \kagome
lattice antiferromagnet.}
\end{center}
\vskip 1cm
\begin{center}
R.Shankar \footnote 
{email: shankar@imsc.ernet.in} $^{\dagger}$
and D.Shubashree
\footnote {email: shuba@mri.ernet.in}.
 $^{\dagger \dagger}$ 

$^{\dagger}$ Institute of Mathematical Sciences, 
Tharamani, Chennai, INDIA, 600113.\\
$^{\dagger \dagger}$ Mehta Research Institute, Chhatnag road, 
Jhusi, Allahabad, INDIA, 211019. \\
\end{center}
\begin{abstract}
In this paper, we study the phases of the Heisenberg 
model on the \kagome  lattice with antiferromagnetic nearest neighbour coupling $J_1$ and ferromagnetic 
next neighbour coupling $J_2$. Analysing the long 
wavelength, low energy effective action that describes
this model, we arrive at the phase diagram as a function of
$\chi = \frac{J_2}{J_1} $. The interesting part 
of this phase diagram is that for small $\chi$,
 which 
includes $\chi =0$, there is a phase with no long 
range 
spin order and  with gapless and spin zero  low lying 
excitations. We discuss our results in the context
of earlier, numerical and experimental work.

\end{abstract}
\newpage
\section{Introduction}
The Heisenberg antiferromagnet (HAF) in two dimensions has 
been widely studied in the last decade from several 
viewpoints. One main motivation for this study has been 
the possibility of encountering novel ordered and disordered
groundstates.  The nearest neighbour HAF on the 
 \kagome lattice (NNKLAF) 
is one system which is expected to show such 
interesting behaviour. This model has been studied
experimentally,  \cite {town}- \cite{orb}  and 
theoretically, \cite{chub} - \cite{thesis}, through 
several methods and is expected to have
a spin disordered ground state. An interesting,
recent study is an exact spectra analysis 
\cite {phil}
 of finite sized $ J_1 - J_2$ model on the \kagome
lattice shows
that as $J_2 \rightarrow 0$, 
there is a spin disordered groundstate  with a gap for excitations 
with spin. This gap is filled with a large number of closely 
spaced singlet excitations which could collapse to the ground state 
in the limit of the system size tending to infinity.

An approach to the problem of two dimensional 
antiferromagnets which has yielded very good 
results is through  a field theoretical sigma model description 
such as has been developed in \cite{halperin,domb,azaria}. 
 The sigma
model describes the large amplitude, long wavelength fluctuations of
the spin system. It is therefore capable of modelling the ordered and
the disordered phases of the system. Its validity 
only requires the existence of 
short range spin order.
The application of this method to develop the field 
theory for the \kagome antiferromagnet has 
been described in 
\cite {first,thesis}. This is the method that we use
here  to determine the phases of the $J_1 - J_2
$ model on the \kagome lattice and in particular
to understand the ground state and low energy 
spectrum of the NNKLAF.  

Among the several families of Kagom\'e lattice antiferromagnets studied
experimentally are the jarosites and the magnetoplumbite like compound
$SrCr_{8-x}Ga_{4+x}O_{19}$. In these compounds 
additional, next to nearest 
neighbour and interplanar, couplings seem to 
stabilise one or the other of the planar states.
Because of this the compounds fall into two groups.
In the iron jarosites, $KFe_3(OH)_6 (XO_4)_2$ 
with $X=S$ or $Cr$ \cite{town} and in $KCr_3(OD)_6(SO_4)_2$ \cite{lee} which
realise $S={5\over 2}$ and $S={3\over 2}$ \kagome
lattice antiferromagnets respectively, ${\bf q}=0$ 
long range order has been observed. In the other group made up of
deuteronium jarosite, 
$D_3OFe_3(OH)_6(SO_4)_2$ ($S={5\over 2}$ ) \cite{wills} and  
$SrCr_{8-x}Ga_{4+x}O_19$ ($S={3 \over 2}$ ) \cite{orb}, 
short ranged $\sqrt 3 \times \sqrt 3$ order is found. In addition it is observed in
these two compounds that the low temperature specific heat 
has a $T^2$ behaviour. 
The usual interpretation of such a behaviour is that there are gapless excitations in the low energy spectrum. 
Such a gapless excitation is usually a Goldstone 
mode 
resulting from the breaking of some continuous 
symmetry in the model. In this case the symmetry 
that can be broken is the $SO(3)$ spin rotational 
symmetry of the hamiltonian. 
However the neutron scattering experiments show that 
there is only short ranged $\sqrt 3 \times \sqrt 3 $
order in the groundstate thereby negating this possibility.
Therefore there is no direct explanation for the 
low temperature specific heat data.

There is an explanation for this puzzle, which we have explored in 
\cite{first}, where we give the mechanism for getting a gapless mode
even in systems where all the symmetries of the microscopic Hamiltonian
are intact and there is no long range antiferromagnetic order.Namely that the low energy theory acquires
an extra continuous symmetry which is not present 
in the microscopic model and this symmetry breaks, 
giving rise to a gapless boson. This is what we call 
a hidden Goldstone mechanism. We work 
out this mechanism in detail for the $J_1 - J_2$ model described below
and analyse the phases as the ratio, $\chi \equiv \frac{J_2}{J_1}$, is
varied.

 This model is defined by the Hamiltonian,

\begin{equation}
H~=~J_1\sum_{\langle ij \rangle}\vec S_i.\vec S_j~-~
 J_2\sum_{\{ ij \}}\vec S_i.\vec S_j
\label{ham}
\end{equation}
Where $<i,j>$ implies that i and j belong to neighbouring 
sites and $\{i,j\}$ implies that i and j belong are next to nearest 
neighbours. 
For positive value of the next neighbour coupling 
$J_2$ the $\sqrt 3 \times \sqrt 3 $ state is picked out from among the 
numerous degenerate groundstates of the classical NNKLAF. 
We write down a field theory 
in terms of the 
five relevant fields, which we identify from a preliminary spin wave analysis
of this model. This theory is an improvement on 
the spin wave analysis since this allows for large 
amplitude fluctuations of these five parameters. 
The theory is symmetric under $SO(3)_R \times SO(2)_L$,
 where the $SO(3)_R$ is the spin rotation symmetry of the 
Hamiltonian
and the $SO(2)_L$ is a special symmetry of the effective low energy action.
We analyse this field theory using the large N expansion described 
in \cite {polyabk} and identify 
that there are two phase transitions as we move towards $J_2 =0$, which 
is the case of the NNKLAF. With reference to figure
(\ref{phases}), for large $\chi$,
we find that the system is in the 
planar spiral phase, the ground state is the $\sqrt 3 \times \sqrt 
3$ state and the low lying excitations about this state are the three 
gapless, magnons. In terms of the field theory this involves the 
symmetry 
breaking pattern  $SO(3)_R \times SO(2)_L \rightarrow SO(2)_L $. 
Reducing $J_2$,  thereby reducing $\chi$, takes us into a 
completely disordered phase where all the symmetries are intact. Further 
reducing $J_2$ takes us into the non-coplanar phase where the symmetry breaking pattern
is $SO(3)_R \times SO(2)_L \rightarrow SO(3)_R $. In this phase, since 
the $SO(3)_R$ symmetry is intact, all correlations of vector and tensor operators 
constructed out of the spins are short ranged and because of the 
breaking of the $SO(2)_L$ symmetry, there is one gapless, spin singlet goldstone mode. 
For reasons that will be clear later we call this the non-coplanar phase
At the lattice level this $SO(2)_L$ symmetry manifests itself as a 
discrete symmetry 
of rotation by $2\pi/3$ followed by a translation by one lattice vector
and at the level of the low energy long wavelength effective action
 this gets enhanced to a continuous
symmetry. The theme of this paper is how this hidden Goldstone mechanism 
gives an explanation of the behaviour of the group two compounds.

The plan of the paper is as follows. In section (\ref{secgs}), we describe the $\vec K =0 $ spinwave 
analysis and 
isolate the five relevant low energy modes. In 
section (\ref{secparam}), we extend the description 
of these low lying modes to include large fluctuations so that the five parameters regroup into a $SU(2) $
matrix valued field and a unit vector field. In section 
(\ref {secop}), we describe the transformation of these fields under the symmetry operations described above. 
In section(\ref{secftandphases}) we describe the 
effective field theory which describes the model 
and has been derived in \cite{first}
and show that there are three distinct phases in the 
$J_1 - J_2$ \kagome lattice model. The details of these phases, in particular the non-coplanar
phase at  $\chi =0$ are described in section 
(\ref{secphases}). Section (\ref{seccorr}) contains 
details of the behaviour of the correlation functions in this 
important phase and section (\ref{secconclu}) contains a 
discussion of our results in the context of 
earlier numerical studies.
\section{Ground state and low energy modes}
\label{secgs}
\begin{figure}
\centering
\hspace{-2cm}
\epsffile{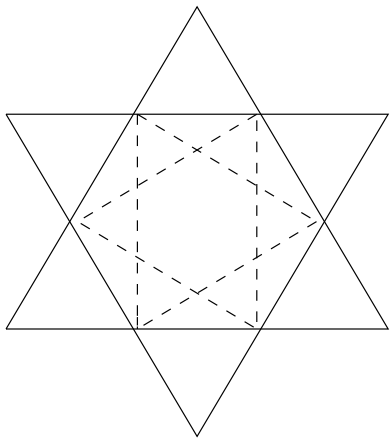}
\caption{In the figure the dashed lines refer to bonds of strength
$J_2 = \stackrel{~}{\chi} J_1$ and the bold lines to bonds of strength $J_1$.}
\label{bonds}
\end {figure}

\begin{figure}
\centering
\hspace{-2cm}
\epsfxsize = 8cm
\epsffile{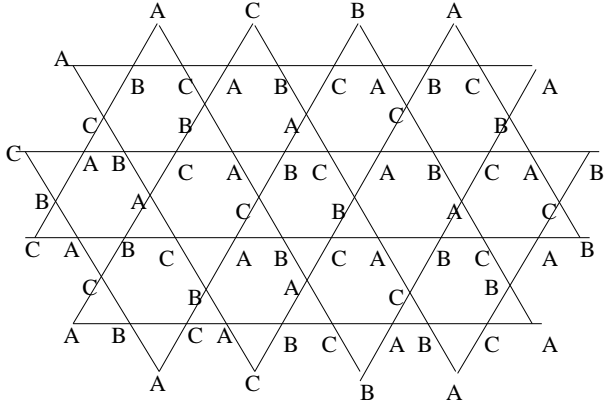}
\caption{The figure shows the $\sqrt 3 \times \sqrt 3 $ state 
superposed on the Kagome lattice , where one choice
for A,B and C is $\vec S_A =(1,0,0), \vec S_B 
= (\frac{-1}{2}, \frac{\sqrt 3}{2}, 0), \vec S_C
=(\frac{-1}{2}, \frac{-\sqrt{3}}{2}, 0) $.
} 
\label{rootthree}
\end {figure}

The model that we consider in this paper is the Heisenberg hamiltonian on the Kagom\'e
lattice with nearest neighbour antiferromagnetic coupling with strength $J_1$ and 
next nearest neighbour couplings with strength $
J_2 = \chi J_1$. This is the model defined in equation (\ref {ham}).
The two types
of bonds are illustrated in figure ({\ref
{bonds}}).
The hamiltonian 
can be rewritten (upto additive constants) as,
\begin{equation}
{H \over J_1}~=~{1 \over 2}\sum_{\Delta}(\sum_{it}\vec S_{it})^2
~-~\chi {1 \over 2}\sum_{\Delta^\prime}(\sum_{it^\prime}\vec S_{it^\prime})^2
\label{ham1}
\end{equation}
where, $\Delta$ denotes the nearest neighbour triangles and $it$ their three vertices.
$\Delta^\prime$ denotes the next nearest neighbour triangles that lie in the hexagons
and $it^\prime$ their vertices. It is clear from equation (\ref{ham1}) that
 for all $\chi > 0$, the
energy is minimised when the magnetisation of all the nearest neighbour triangles is
zero and that of all the next nearest neighbour triangles is the maximum possible. This
occurs for the $\sqrt 3 \times \sqrt 3$ state shown in figure ({\ref{rootthree}}).

   In order to identify the lowest
lying excitations about this ground state, we
do a spinwave analysis. In this analysis we
treat the quantum fluctuations as rotations
of the classical spin vectors, thereby mapping
the spin variables $\vec S_j $
on to bosonic variables $P_j $ and $Q_j$. 

In order to do this we rewrite the spins $\vec S_j$ as,
\beq 
\vec {S_j} = e^{\frac { -i w_j ^a T^a}{\sqrt
S}} {\vec S_j
^{cl}}
\label{spin1}\eeq
where the $T^a$ are the three generators of $SU(2)$
in the spin 1 representation. The 
$w_j^a = P_j E_j ^a +Q_j Z^a $ , $P_j$ 
and $Q_j$ being bosonic operators obeying the 
commutation rules $[P_j, Q_{j'}] = -i \hbar \delta _
{j,j'}$.
${\vec S_j^{cl}} = S \hat n_j$, where the  $\hat n
_j$ are arranged according to the $\sqrt 3
\times \sqrt 3 $ configuration and the set
$\{\hat n_j , \hat E_j , \hat Z \}$ form an
orthogonal triad at each site. 
Thereafter, for small fluctuations, putting these definitions 
into equation (\ref{spin1}) and expanding to order $\frac{1}{S}$
, we get,
\beq
\vec S_j = S\hat n_j ( 1 - \frac{P_j^2 + Q_j^2}
{2S}) + \sqrt S \hat E_j P_j + \sqrt S \hat Z
Q_j 
\label{spin2}\eeq 

\begin{figure}
\centering
\hspace{-2cm}
\epsfxsize = 7.5cm
\epsffile{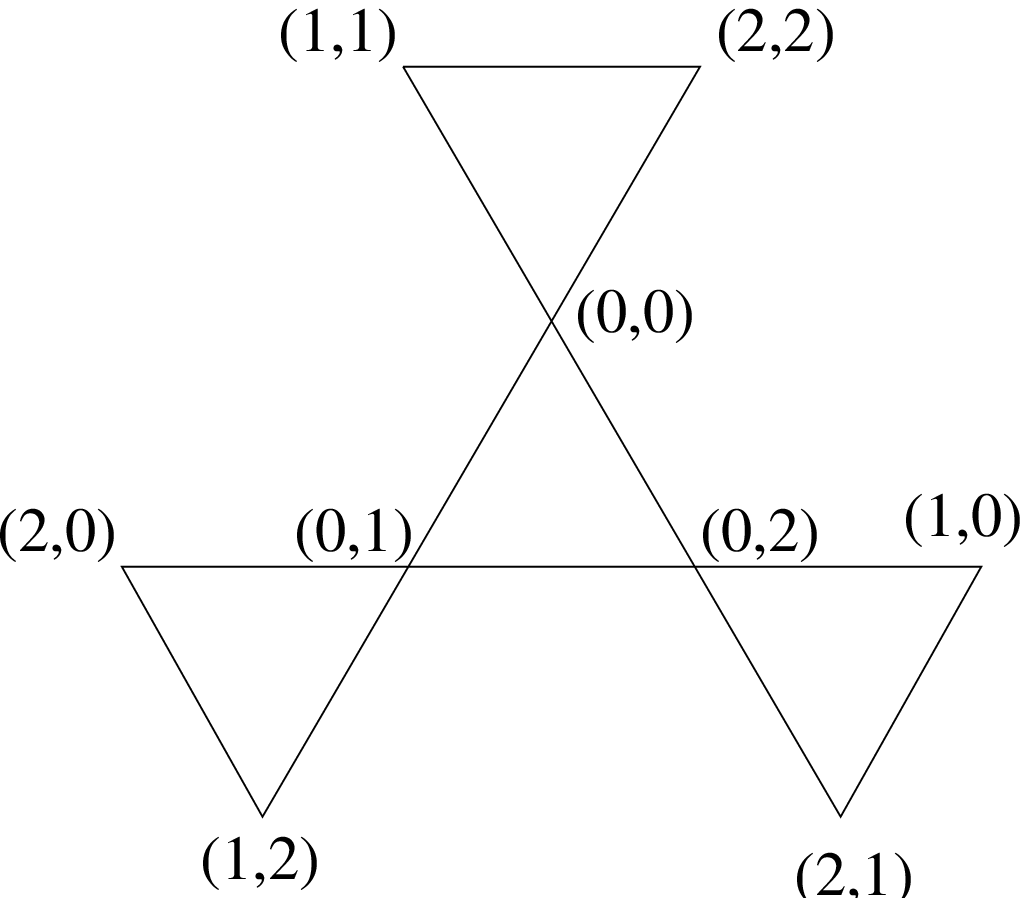}
\caption{ }
\label{ucell2}
\end {figure}

Before proceeding, we notice that 
the lattice splits into magnetic unit cells consisting 
of 9 points each
because of the periodicity of the lattice 
and the $\sqrt 3 \times \sqrt 3 $ groundstate
. Therefore we expand our
notation a bit, and replace  $\vec S_j $ 
equivalently by  $ \vec S_{Jj\beta}$. In this 
notation every lattice index j is replaced 
by one unit cell index such as J and  two 
sublattice indices $(j,\beta)$. This labelling
is shown in figure (\ref {ucell2}) which also 
shows the structure of each unit cell.
We make such an expansion of $S_{Jj\beta}$ 
as in equation (\ref{spin2}) , keeping upto 
quadratic terms in the P and Q. We finally
get  the fluctuations hamiltonian in terms 
of the Fourier transformed variables $P_{Kj\beta}$ and $Q_{Kj\beta}$, which are defined as
follows.
\beqar
P_{Kj\beta}&=& \frac{1}{N} \sum _{J} P_{Jj\beta}
e ^{-i {\vec X_J}.{\vec K}} \nonumber \\
Q_{Kj\beta}&=& \frac{1}{N} \sum _{J} Q_{Jj\beta}
e ^{-i {\vec X_J}.{\vec K}} \nonumber \\
\eeqar
In terms of the ${\bf P_K }$ and ${\bf Q_K}$ the 
hamiltonian reduces to a $9 \times 9 $ block
which is given by,
\begin{equation}
H_{sw}~=~{1 \over 2} \sum _{\bf K}
 {\bf P_{-K}^T}M^{-1}
{\bf P_{K}}~+~{1 \over 2} {\bf Q_{-K}^T}K
{\bf Q_{K}}
\label{sw0}
\end{equation}

The eigenvalues and the eigenvectors 
of these
matrix $\Omega ^2 = M^{-1} K$ for $\vec K = 0$ give the gaps and
normal modes of the system of oscillators. 
These  matrices are given in 
the appendix (A). The
9 eigenvectors of $\Omega ^2 $, denoted by $\phi _{j \beta}$
 are given by $\phi_{j \beta} = e_j \times
e_{\beta}$, 
where, 
\beqar
 e_0 = \frac{1}{\sqrt 3} (1,1,1) \nonumber \\
 e_1 = \frac{1}{\sqrt 3} (1,\alpha,\alpha ^2)
\nonumber \\
 e_2 = \frac{1}{\sqrt 3} (1,\alpha^2,\alpha)
\eeqar
The eigenvalues corresponding to the eigenvectors $\phi
_{j\beta}$ are $\omega^2_{j\beta}$. These fall into three 
groups. 
$\omega ^2 _{00} = \omega^2_{01} = \omega^2_{02} = 0$,
$\omega^2_{11} = \omega^2_{22} = 18\chi (3\chi +1)$ and
$\omega^2_{10} = \omega^2_{12} = \omega^2_{20} = \omega^2_{21}
= 36 \chi ^2 + 25 \chi + 9/2$.  Accordingly, we classify 
the first three modes as S-S modes since they are soft 
for all $\chi$, the second pair as H-S modes since they 
are gapless for $\chi = 0$ and hard for $\chi \not = 0 $
and the last four as the H-H modes as they are hard for 
all $\chi$. In the large $\chi $ regime the S-S modes are the 
only relevant modes, while we need to take both the S-S modes 
and the H-S modes into consideration to describe low energy 
physics in the small $\chi $ domain.

\section{Parametrisation of large amplitude
fluctuations.}
\label{secparam}
In this section, starting from the expression for the normal 
modes of the $\vec K =0$ spinwave hamiltonian, we develop 
a parametrisation of the S-S and H-S modes for large amplitude
fluctuations. These five modes, since they are gapless at the 
NNKLAF end, will govern the  low energy physics, both at the NNKLAF 
end and close to it, i.e for small  $\chi$ . 
For $\chi =0$, some of these modes are
 dispersionless. This arises from the possibility of having local 
fluctuations about line defects and have been discussed in 
\cite {chalk}.

From the expression in equation (\ref{spin1}) for the spins, 
putting in the forms of the eigenmodes for the five relevant 
modes as expressions for the $P_{j \beta} $ and 
$Q_{j\beta}$, we observe that for $\vec K =0$, all 
the modes satisfy $
\vec S_1 + \vec S_2 + \vec S_3 = \Delta \vec S = 0$, for the 
spins $\vec S_1 , \vec S_2, \vec S_3$ lying on a triangular plaquette.
For the dispersionless H-S modes, this identity continues to 
hold for $\vec K \not = 0$.

At the spin wave level, the expansion of  $ \vec S_{j \beta}$
to order $\frac{1}{S}$ indicates that  we are looking at 
small fluctuations about the classical ordered configuration. 
For the purposes of the field theory we need to parametrise large amplitude fluctuations of these relevant modes. We now 
proceed to do this.

Though an 
exact treatment would involve deriving this 
for $\vec K \not = 0$, we use the zero $\vec K$
expressions as an approximation. This is valid
because we are interested only in the long 
wavelength excitations. 

If the 9 spins of the unit cell may be thought of
as a rigid unit, the three S-S modes correspond
to the rotation of this unit about the three 
co-ordinate axes. This set of three modes can 
therefore be parametrised by a unitary matrix
$U_J$, which brings about this rotation from 
the ground state configuration to the body fixed
frame of the rigid body. This is the usual 
interpretation given to the gapless Goldstone modes
that occur in antiferromagnetic models. 

Since the H-S modes cost zero energy at $\chi =0$,
 they must leave the relative angles between the 
spins on each triangle intact. They could however,
distort the angles between neighbouring triangles.
Hence they create
non planar configurations within the unit cell. 
We can parameterise these configurations as follows,
\begin{equation}
\vec S _{Jj\beta} = V^{\dagger}_{Jj} \hat n _{\beta}
\label{hsparam0}
\end{equation}
$V_{Jj}$ are rotation matrices that rotate the three
spins (labelled by $\beta$) belonging to each 
$\sqrt 7 \times \sqrt 7 $ triangle
(labelled by the same $j$) rigidly. They are not independent of 
each other but are constrained by the fact that the 
inter-unit cell triangles are also not distorted. It is
difficult to solve for these constraints exactly. However,
we can use the eigenfunctions of the H-S modes to obtain
the following, approximate, solution.
\begin{equation}
V_{Jj} = \exp ^{i\frac{2\pi}{3}j T^{3}}\exp ^{i
\frac{\pi}{3}\hat m _J . \vec T}\exp ^{-i\frac{2\pi}{3}j T^{3}} \exp ^{-i\frac{\pi}{3} T^3}
\label{hsparam}\end{equation}
When $\hat m = \hat z$, $V_{Jj}$ reduces to the identity
matrix and the spin configuration is undistorted. When 
$\hat m$ deviates from the z-axis, we can show that the
configurations produced by equations (\ref{hsparam0})
 and (\ref{hsparam}) are,
upto quadratic order in the deviations, zero energy configurations
in the $\chi \rightarrow 0$ limit. We therefore use equations
(\ref{hsparam0}) and (\ref{hsparam}) to approximate 
the long wavelength, 
large amplitude fluctuations of the H-S modes.

If the spins on the unit cell form a rigid unit and
if the ground state configuration is thought of as
a space fixed frame, then the S-S modes,
parametrised by the $U_J$, rotate the spins to the
body fixed frame. further the $V_{Jj}$ cause a
rotation in this body fixed frame, whose magnitude
and direction are decided by $\hat m$ which is 
a vector in this body fixed frame. $\vec S_{Jj\beta}$ is 
therefore given by,
\begin{equation}
\vec S _{Jj\beta} = U^{\dagger} _{J} V^{\dagger}
_{Jj} \hat n _{\beta}
\end{equation}

\section{Order parameters and symmetries.}
\label{secop}

Now that we have a  parametrisation of the large 
amplitude fluctuations of the five relevant modes,
it is necessary to look at the symmetries of the 
microscopic model and how they act on the fields
$U_J$ and $\hat m _J$. There are two important
groups of transformations which leave the hamiltonian
 invariant. The first is a global $SO(3)$ group of 
rotations  of each spin under the transformation 
\beq
S^a _{Jj\beta} \rightarrow \Omega
 _R  ^{\dagger ab} S_{Jj\beta} ^{b}
\eeq

Since this is an overall rotation of the spins
with respect to the space fixed frame, this simply
adds on to the matrix $U_J$ and leaves $\hat m_J$
invariant, thus,
\beqar
U_J  \rightarrow U_J \Omega _R \nonumber \\
\hat m _J \rightarrow \hat m _J
\eeqar

Secondly, the hamiltonian is invariant under the 
rotation of each spin by $\frac{2\pi}{3}$ followed
by a unit translation.Under this transformation the 
spins transform as follows,
\beq
S^a_{Jj\beta} \rightarrow R^{\beta \beta'} S^a_{Jj\beta'}
\eeq
 In the continuum limit
this shows up as a combination of $U(1)$ 
rotations in both the space fixed frame and in 
the body fixed frame, under which the fields transform as 
follows,

\beqar
U_J \rightarrow \Omega _L U _J \nonumber \\
\hat m_J \rightarrow \Omega _L \hat m_J
\eeqar
Looking at the way these rotations act on $U_J$, 
we call them $SO(3)_R $ and $SO(2)_L$ rotations
respectively.

As mentioned in the previous section, the H-S modes
create non-coplanar configurations in the NNKLAF. This
is contrary to the effect of the S-S modes which 
 are rigid rotations of the spins of the unit cell. 
Hence the  effect of the H-S modes may be measured 
by defining an order parameter that leaves the 
spins on the triangular plaquettes rigid, yet 
distorting the planarity of adjoining triangles. 
A suitable candidate is the scalar triple product
of three spins lying in a row on each unit cell. 

We show in section (\ref{seccorr}) which follows, 
that  this is a suitable order parameter with 
which to describe the phases of the NNKLAF.

\section {Field theory and phases.}
\label{secftandphases}
In our previous work, \cite{first}, we have derived
the long wavelength, low energy field theory describing 
the deformed triangular antiferromagnet close to the 
Kagom\'e lattice limit. Based on the same considerations
the same field theory would also describe 
the $J_1 - J_2 $ model on the Kagom\'e lattice for 
$J_2 $ close to zero. This action is given by the expression,

\begin{equation}
S[\phi^a_{~r},\hat m] = \int d^3x~\partial_\mu\phi^a_{~r}\partial_\mu\phi^a_{~r}
                       ~+~{1 \over g_2}\partial_\mu\hat m.\partial_\mu\hat m
                       ~+~V(m^z)
\label{action1}
\end{equation}
Along with the constraint,
\begin{equation}
\sum _{r}\phi^a_{~r}\phi^b_{~r}~=~{1 \over g_1}(1+f(m^z))\delta_{ab}
\label{cons}
\end{equation}
The interaction is built into the constraint which 
puts $\phi ^a _r = \sqrt {\frac{1}{g_1} (1 + f(m^z))}
 \Phi _r ^a $,
where the fields $\Phi _r ^a $ make up the columns of the matrix U
and the function $f(m^z) = \alpha (1 - (m^z)^2)$. 
The potential $V(m^z) = \lambda _0 ((m^z)^2 - \eta _0) ^2$. 
The parameters $\alpha , \lambda _0 , \eta _0 $ and 
g are all functions of $\chi$.

In \cite{first}, we had shown that when $g_1$ is in the strong
coupling regime and $g_2$ in the weak coupling regime, there exists a phase
where the $SO(3)_R$ symmetry is unbroken and the $SO(2)_L$ is broken. We now give
a simple large N formalism where the physics of this regime can be analysed in
a systematic 1/N expansion. 

The large N formalism we use is of the standard type used to analyse disordered
phases of non-linear $\sigma$ models \cite{polyabk}. $\phi^a_{~r}$ can be thought 
of as a set of three orthogonal 3 dimensional vectors. This is generalised to 
a set of three orthogonal N dimensional vectors. We denote them by $\phi^a_{~r}$, 
where $a~=~1,2,\dots ,  N$ and $r~=~1,2,3$. The coupling constants in the 
model are defined to scale with N as follows. $g_{1(2)} \rightarrow g_{1(2)}/N,
~\alpha \rightarrow N\alpha$ and $\lambda_0 \rightarrow N\lambda_0$. This results
in the RHS of the constraint in equation(\ref{cons}) to be multiplied by N. We then
use a $3 \times 3$ matrix valued Lagrange multiplier field, $\mu^{ij}(x)$, to
impose the constraint as is usual in this method. The $\phi$ fields can then be 
integrated out and the partition function is expressed as,
\begin{equation}
Z~=~\int_{\mu,\hat m}e^{-NS_{eff}[\mu,\hat m]}
\label{part}
\end{equation}
where, $S_{eff}[\mu, \hat m]$ is given by,
\begin{equation}
S_{eff}[\mu,\hat m]={1 \over 2}lndet(-\partial_\mu^2-i\mu)
                              +\int_x~{1 \over g_1}(1+f(m^z))tr\mu
                              +{1 \over g_2}(\partial_\mu \hat m)^2 
                              +V(m^z)  
\label{seff}
\end{equation}
It is now clear that using the saddle point method, a systematic expansion
in 1/N can be developed for $Z$. The same procedure can easily be generalised 
for correlation functions as well.

The saddle point equations are obtained by setting the variation of the effective 
action in equation(\ref{seff}) with respect to $\mu^{ij}(x)$ and $\hat m(x)$ equal 
to zero. We look for translationally invariant ($x$ independent) solutions. Putting 
$m^z = cos(\theta)$ and $-i\mu^{ij} = M_a^2 \delta^{ij}$, the saddle point
equations can be written as, 
\beqar
\frac{\delta S}{\delta \mu ^{ij}}~~&=& 0 \label{stcond1} \\
\frac{\delta S }{\delta (m^z)^2}~~&=& 0 \label{stcond2}
\eeqar
Applying these conditions and 
putting in for $\mu _{ij}$ the ansatz,
$i\mu _{ij} = M_a ^2 \delta _{ij}$ we get the following solutions
for $m^z$,
\beq
\frac{1}{g_1} - \alpha (m^z)^2 = \frac{1}{2}\int \frac{d^3 k}{(2\pi)^3}
\frac{1}{K^2 + M_a ^2} 
\label{gapeqn0}\eeq 
From the second condition (\ref{stcond2}), 
we get the two possible solutions for 
$m^z$  of which the first one is,
\beq
m^z = \eta _0 - \frac{2 \alpha M_a ^2}{\lambda _0} = \cos \bar \theta
\label{theta}
\eeq
The other possible condition solution is  $m^z = 1$. 

If we define $g_{crit}$ by the equation,
 
\beq
\frac{1}{g _{crit}} = \int \frac{d^3K}{(2\pi)^3} \frac{1}{k^2}
= \frac{\Lambda}{2\pi^2} 
\eeq
Then when 
\beq
\frac{1}{g_1} - \alpha \geq \frac{1}{g_{crit}}
\label{cond1}\eeq 
Then, $M_a^2 =0 $ and from equation (\ref{theta}) , 
if $\eta_0 >1$, the solution for $m^z$ 
in this case is $ m^z =1$.
Since in this phase $\frac{1}{M_a}$, which is the 
correlation length for the $\Phi $ fields, diverges,
this describes	
an $SO(3)_R$ broken phase. In addition, in this phase 
 the $SO(2)_L$ symmetry is unbroken because
 $m^z =1 $.
The 
low lying excitations in this phase are the three 
Goldstone bosons coming from this symmetry breaking. 
They are the three spinwave modes with spin = 1.

When 
\beq
\frac{1}{g_1} - \alpha < \frac{1}{g_{crit}}
\label{cond2a}\eeq
Then 
$ M_a \not = 0 $ ( which follows from (\ref{gapeqn0})
)
  and this is a phase in which 
the  $SO(3)_R$ symmetry is unbroken.
In this disordered regime, the fields $\Phi $  have a finite 
correlation length which can be calculated 
by solving equation (\ref{gapeqn0}) to get $M_a$. 

Now there are two possible solutions for $m^z$. 
If 
\beq
\eta_0 - \frac{\alpha M_a^2}{\lambda _0} > 1 
\label{cond2b}\eeq
Then this offers no solution to equation 
(\ref {theta})for $m^z = \cos \bar \theta$ which 
is always less
that or equal to 1. Hence the other solution, $m^z =1$ is picked out
 . This implies that this is also a  phase 
in which the  $SO(2)_L$ 
symmetry is unbroken. The groundstate , in this phase 
shows no long range order and all low lying excitations about 
this state are gapped.

The third possibility is when 
\beq
\eta_0 - \frac{\alpha M_a^2}{\lambda _0} < 1
\label{cond3}\eeq
In this case there is a  consistent alternate
solution to equation (\ref {theta}) for 
$m^z = \eta _0 -\frac{\alpha M_a^2}{\lambda _0} =
\cos \bar \theta$. Since $\hat m$ no longer points 
in the $\hat z $ direction, this spoils the axial 
symmetry that existed earlier. In this phase, the $SO(3)_R$ symmetry is, as
earlier, unbroken but the $SO(2)_L$ symmetry is broken and there
is one massless particle which is the angular variable 
$\phi _m$. Since $\hat m$ is a spin singlet under $SO(3)_R$
rotations this is a spinless excitation. The other field $\theta _m$ acquires
a gap,which can be calculated.

In this $SO(2)_L$ broken phase the fluctuations in $\theta _m$  are  gapped and 
those of $\phi _m $ are gapless. We obtain the mass gap for the $\theta _m$ 
fluctuations and 
rewrite the part of the action S$_m$, which involves just the
fields $\hat m$, in terms of the variables $\theta 
_m $ and $\phi _m$ we get, up to quadratic order in the fields,
\beq
S_m = \int d^3 x \sin^2(\bar \theta _m) 
\partial _\mu \phi _m \partial _\mu \phi _m
 + \partial _\mu \theta _m\partial _\mu \theta _m +
\frac{{M_\theta} ^2}{2} (\theta _m)^2
 \label {Sm-theta-phi}\eeq
Where $M_{\theta} $ is got by  
expanding $S_m$ about the average value of  
$m_3 = \cos \bar{\theta}$  
given in equation (\ref{theta}). 
This is given by the expression,
\beq
M_\theta ^2 = \frac{\lambda _0}{g_2}[ (2 \eta _{0}  - 3) \cos (2 \bar \theta)
- \cos (4  \bar \theta )]
\eeq

So far the discussion  has been restricted to a 
regime where the $\hat m$ fields are approximated 
by their classical values. 
We still need to ascertain 
that including fluctuations in $\hat m$ does not 
destroy the ordered state.
This has been  established in \cite{first} within 
a one loop R.G calculation. 
This calculation showed that while the fluctuations 
of $\hat m $  tend to destroy the order, there 
exists a region of parameter space where they do 
not succeed in doing so. This ensures the stability 
of the 
$SO(2)_L$ broken phase over the effect of quantum
 fluctuations.
\section{Phases of the Kagom\'e lattice.}
\label{secphases}

\begin{figure}
\centering
\hspace{-2cm}
\epsfxsize = 10cm
\epsffile{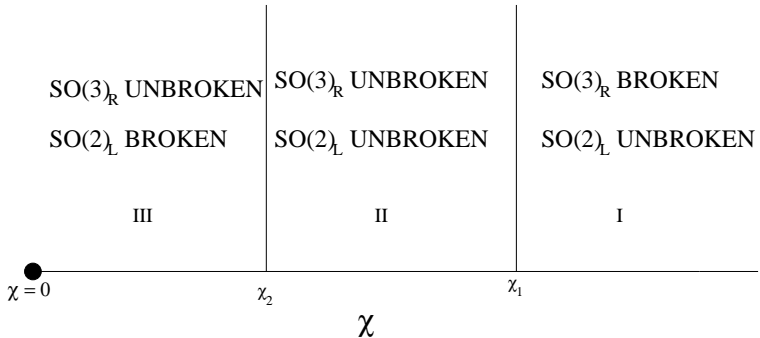}
\caption{phase diagram  of the $J_1
- J_2$ model on the \kagome lattice. Here 
$\chi = \frac {J_2}{J_1}$ and the point 
$\chi =0$ is the NNKLAF. }
\label{phases}
\end {figure} 
In the previous section we have analysed the 
longwavelength field theory (\ref{action1}) and shown that 
the system  undergoes two phase transitions as 
the coupling constants are tuned.
To relate these results to the phases of \kagome lattice 
model that we are considering, we need to relate the coupling
constant of the lattice model $\chi$ to the coupling constants
of the field theory. It is very difficult to reliably calculate
this relation. However, based on some general features and the 
numerical results obtained by Lecheminant et. al. \cite{phil}, 
it is possible to make some qualitative statements.

Firstly it is possible to estimate the potential, $V(m_z)$ and
hence the parameters $\eta_0$ and $\lambda_0$ by substituting
equation (\ref{hsparam}) into the hamiltonian and taking $\hat m_I$
to be independent of $I$. By doing so we get, $\eta_0=1 + 4 \chi $ and 
$\lambda_0=\frac{27}{4}J_1 $ \cite{thesis}. While this will be modified by the fluctuations, we
assume that the qualitative fact that $\eta_0$ is an increasing function
of $\chi$ and approaches $\eta_O=1$ as $\chi \rightarrow 0$, remains true. 
This amounts to assuming that the gap of the singlet excitations (the H-S
modes) decreases as $\chi$ decreases and approaches 0 as $\chi 
\rightarrow 0$. This is consistent with our spinwave spectrum and the 
results in reference \cite{phil}

Next, as $\chi$ decreases and the corresponding bonds become weaker, we
can expect large amplitude fluctuations of the spins to cost less energy
and therefore for the spins to disorder. The results of reference \cite{phil}
strongly support this scenario. We therefore assume that $g_1$ increases as
$\chi$ decreases. 

Now we look at equation 
(\ref{cond2a}) which determines the first phase
boundary for the transition from phase I to phase II (see fig.(\ref{phases})).
The above assumptions imply that for large $\chi$ the system is in phase I 
and as $\chi$ is decreased and we move from phase I 
to phase II the correlations of the $\Phi$ fields 
become short ranged and the corresponding mass gap
$M_a$ increases from the value zero (at $\chi _1$).
$\eta _0$ takes the value one for the NNKLAF and 
increases as $\chi$ is increased.
As $\chi$ is 
further decreased, $M_a$ increases and 
 from equation (\ref{cond3}), for the second phase 
boundary, we see that 
at some point the inequality is 
saturated. This is the second critical point $\chi_2$. 

This forms the argument for the scenario depicted in figure (\ref{phases}). 
For $\chi >\chi _1$, 
the system is in the N\'eel ordered spiral phase.In this 
phase, marked I in the diagram,
the $SO(3)_R$ symmetry is broken down to nothing 
and the $SO(2)_L$ symmetry is unbroken. The value 
of $\chi _1$ is determined by the saturation of the 
inequality  (\ref{cond1}). 
As $\chi $ is reduced below $\chi _1$ the system 
undergoes a transition into the phase marked II. 
This is a phase with all symmetries intact . 
Further down, there is a second phase transition at 
$\chi = \chi _2$, which is determined by the condition (\ref{cond3}).
The phase III is one in which the $SO(3) _R $ symmetry 
continues to be unbroken and the $SO(2)_L$ symmetry 
is broken. 

It has been seen in reference \cite {phil}
that for  $J_2 =0$ the system is disordered and a large number 
of low lying singlet 
excitations that are observed.  In our analysis of the 
continuum model, the breaking of the $SO(2)_L$ symmetry 
would result in the collapse of such states
onto the ground state as the system size is increased to infinity.
Our analysis predicts an intermediate phase 
II, where both the symmetries are intact. But since it seems
that the ordering of the $\phi _ m$ field is driven by the 
disordering of the $U$ field and vice-versa, 
it is possible that in the real system, fluctuations 
could cause $\chi_1$ and $\chi _2$ to coincide, 
thereby causing a direct jump from phase I to phase III. 

This is our picture of the phases of the 
Kagom\'e lattice as a function of $\chi$. In the next
section we describe the behaviour of the correlation 
functions in the phase III and give an 
expression for the relevant order parameter.

\section{Correlation functions and order parameter.}
\label{seccorr}

We will now construct a suitable local order parameter, in terms of the 
spins, to describe the phase III discussed in the previous section.
This is the phase in which the $SO(3)_R$ spin symmetry is unbroken and the
$SO(2)_L$ symmetry is broken. The order parameter should therefore be a 
spin singlet and transform non-trivially under the $SO(2)_L$ symmetry. In
terms of the field theory variables, the transverse components of the 
$\hat m$ field, i.e $(m_x, m_y)$, is such an order parameter. It will
have a non-zero value in phase III and its correlation functions will be
long ranged. 

\begin{figure}
\centering
\hspace{-2cm}
\epsffile{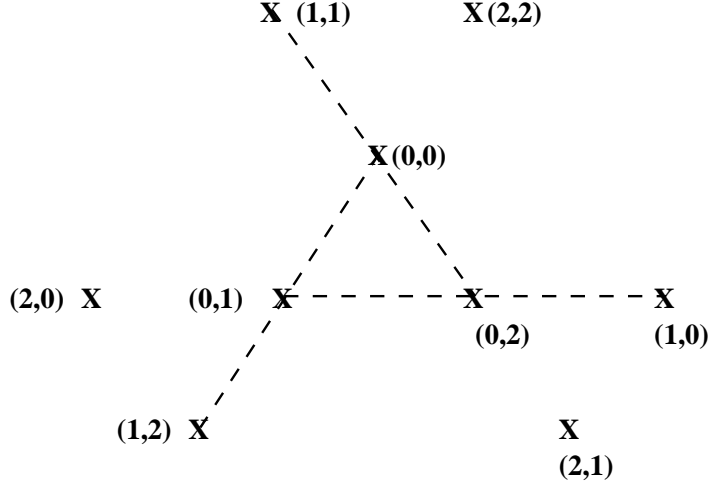}
\caption{The crosses refer to various sites on the unit cell and the dashed lines connect sites forming 
the order parameters C(X,11,02,00), C(X,12,00,01) and C(X,10,01,02) .}
\label{corr}
\end {figure}

Since the $SO(2)_L$ symmetry is not present in the lattice 
spin model, the identification of such an order parameter in terms of the 
spins is not straightforward. As discussed in section(\ref{secparam}), the
$\hat m$ fields represent spin configurations where the elementary
triangles are left intact but neighbouring triangles are not coplanar. 
Now consider the scalar triple product of three spins lying on the same line 
in a unit cell. e.g. labelled by (11), (02) and (00)
(see figure (\ref{corr}) ).
\begin{equation}
C(X,11,02,00) = \vec{S}_{11}. \vec{S}_{02}\times \vec{S}_{00}  
\label{stp}
\end{equation}
This operator is a spin singlet. When the two neighbouring triangles that
they belong to are coplanar, then so are these three spins and consequently
the their scalar triple product is zero. When the triangles are not coplanar,
then neither are the three spins. Since $\vec{S}_{02}\times \vec{S}_{00}$ is 
the vector normal to the plane of the central triangle  
and $\vec {S} _{11}$ is a vector lying in the plane of the neighbouring 
triangle, therefore the scalar triple product defined in equation (\ref {stp}) 
is a measure of the angle between the planes of the two adjoining triangles. 
We can therefore expect it to reflect the behaviour of the $\hat m$ field.

To confirm this we express the right hand side of equation(\ref{stp}) in terms
of the $U$ and $\hat m$ fields using the parameterisation of the spins given in
equations (\ref{hsparam0}) and (\ref{hsparam}). Since it is a spin singlet, it
is independent of $\Phi _r ^a$ and it turns out to be,
\beqar
C(X,11,02,00))&=&\frac{3}{4} Cos \phi _m (X) Sin \theta _m(X) \nonumber \\
&=& \frac{3}{4} \hat m(X) . \hat E _1
\label {cval} 
\eeqar
where $\hat E_1 =(1,0,0)$, a vector in the body fixed frame.
Thus C(X,11,02,00) is  proportional to $m_x$. 
Similiarly we can define the operators 
\beqar
C(X,12,00,01) &=& \frac{3}{4} Cos(\phi _m(X) +\frac{2
\pi}{3} ) Sin \theta _m (X) \nonumber \\
&=& \frac{3}{4} \hat m(X) . \hat E _3
\label{stp2}\eeqar
and 
\beqar
C(X,10,01,02) &=& \frac{3}{4} Cos(\phi _m(X) -\frac{2
\pi}{3} ) Sin \theta _m (X) \nonumber \\
&=& \frac{3}{4} \hat m(X) . \hat E _2
\label{stp3}\eeqar
where $\hat E_2 = (\frac{-1}{2}, \frac{\sqrt 3 }{2},
0)$ and $\hat E_3 = (\frac{-1}{2}, \frac{-\sqrt 3}{2}, 
0)$. By taking other
appropriate linear combinations of these, 
we can construct an operator
which is proportional to $m_y$ and we see that operators of the type written down in equations (\ref{cval}, \ref{stp2} and  \ref{stp3})
are good order parameters to characterise the $SO(2)_L$ broken phase. 

\section{Conclusions and discussion}
\label{secconclu}

In this paper, we have described the 
$J_1 -J_2$ model on the Kagom\'e lattice 
by a field theory of the low energy long wavelength 
excitations. Analysis of this theory shows that there
are three phases in this model. Based on a comparison
with earlier , exact diagonalization studies, we 
relate the coupling constants of the field theory 
to the parameter $\chi$ of the hamiltonian. Thereby,
we depict our results as a phase diagram on the 
$\chi$ axis as shown in figure (\ref{phases}).
Accordingly, the NNKLAF lies in the phase III which 
has been described earlier. In this phase, the ground 
state is disordered and there are massless singlet 
excitations over the ground state. At large and 
small $\chi$ our description of the $J_1 - J_2$ 
model matches with the numerical studies. In addition, at intermediate $\chi$, we see a completely 
disordered phase (II). We have described the transition from phase II to phase III by a suitable singlet 
operator constructed out of the spins. 

Further, the field theoretic approach explains the 
origin of the gapless excitation in the disordered 
phase III, by giving a new mechanism.
This mechanism of obtaining a Goldstone mode 
is likely to be operative in the model for the group 2 
compounds and provides  a means of getting a gapless
bosonic excitation which 
lead to a $T^2$ behaviour of the specific heat. This
 is interesting because all symmetries of the 
microscopic model are apparently intact and hence there 
seems to be no reason for the existence of such a 
gapless mode.

In their exact spectra analysis of the $J_1 - J_2$
model on the Kagom\'e lattice, Lecheminant et al \cite {phil}, see 
a trend that is supportive of the above picture as far as the NNKLAF
is concerned. Namely they see that while there is no long range 
spin order at the $J_2 =0$ end, there is a proliferation
of spin singlet excited states with a small gap which could collapse to the 
ground state in the limit of an infinite lattice. 
Comparing our results to this, it seems likely that 
this collapse is a signal of the breaking 
of the $SO(2)_L$ symmetry in the limit of infinite 
lattice size.
\\[1cm]

\appendix
\label{appendixA}
{\bf  Appendix A.} \\

The fluctuations hamiltonian is given by the expression,
\beq
H = \frac{1}{2} \sum P_{-K} M^{-1}P_{K} + Q_{-K} K Q_{K}
\eeq

where the $M^{-1}_{0}$ and The $K_0$  are given 
by,\\[1cm]
$\begin{array}{ll}
 (M^{-1}_0)_{i\alpha j\beta} =  \left [ 
  \begin{array}{ccc}
  (I_0)_{\alpha \beta} &(I_1)_{\alpha \beta}  & (I_1^T)_{\alpha \beta}  \\ 
   (I_1^T)_{\alpha \beta} & (I_0)_{\alpha \beta} & (I_1)_{\alpha \beta} \\
  (I_1)_{\alpha \beta} & (I_1^T)_{\alpha \beta}  & (I_0)_{\alpha \beta} \\ 
  \end{array} \right ] _{ij} 
~~~ & ~~~ 
\\[.5cm]
\mbox{where,}
\\[.5 cm]
(I_0)_{\alpha \beta} = \left [
  \begin{array}{ccc}
  2(1 - 2\chi)  & 1  &1   \\
  1  &  2(1 - 2\chi) & 1  \\
  1  &1   &  2(1 - 2\chi) \\
  \end{array} \right ]_{\alpha \beta}
~~~ & ~~~
(I_1)_{\alpha \beta} = \left [
  \begin{array}{ccc}
 2 \chi & 0 & 1 \\
 1 & 2 \chi & 0 \\
 0 & 1 & 2 \chi \\
  \end{array} \right ]_{\alpha \beta} 
\end {array} $
\\[.5cm]
and
\\[.5cm]
$
\begin{array}{ll}
  (K_0)_{i \alpha j \beta} = \left [ 
  \begin {array}{ccc}
   ({\bar I} _0)_{\alpha \beta}&({\bar I}_1 )_{\alpha \beta}
& ({\bar I }_1 ^T)_{\alpha \beta}\\
   ({\bar I }_1 ^T)_{\alpha \beta} & ({\bar I} _0 )_{\alpha \beta}
& ({\bar I}_1 )_{\alpha \beta}\\
  ({\bar I}_1)_{\alpha \beta} &({\bar I }_1 ^T )_{\alpha \beta} 
&  ({\bar I} _0 )_{\alpha \beta}\\
  \end{array} \right ] _{ij} 
&  \\[.5cm]
\mbox{where,}
\\[.5cm]

  ({\bar I}_0)_{\alpha \beta} = \left [ 
  \begin {array}{ccc}
  2(1-2\chi)& -\frac {1}{2} & -\frac {1}{2} \\
 -\frac {1}{2} & 2(1-2\chi) & -\frac {1}{2} \\
  -\frac {1}{2} & -\frac {1}{2} & 2(1-2\chi) \\
  \end{array} \right ]_{\alpha \beta}  
& ({\bar I}_1){\alpha \beta} = \left [ 
  \begin{array}{ccc}
  2\chi & 0 & -\frac{1}{2} \\
  -\frac{1}{2} & 2\chi & 0 \\
  0 & -\frac{1}{2} & 2\chi \\
  \end{array} \right ]_{\alpha \beta} \\[1cm]

\end{array}
$ \\[1cm]

\end{document}